\documentclass[journal,twoside,final]{IEEEtran}
\usepackage{graphicx}
\graphicspath{{figures/}}

\RequirePackage{xspace}
\def\KAOS{{\sc Kaos}\xspace}
\begin{document}
\bstctlcite{OPTIONS}
  
\title{A Tracking Fiber Detector based on Silicon Photomultipliers 
	for the \KAOS\ Spectrometer}

\author{S.~{S\'anchez Majos},
	P.~Achenbach, and
        J.~Pochodzalla
\thanks{Manuscript received November 14, 2008. Work supported in part by
  GSI as F+E project MZPOCH.}  
\thanks{The authors are with the Institut f\"ur Kernphysik, Johannes
  Gutenberg-Universit\"at, Mainz, Germany (e-mail: sanchez@kph.uni-mainz.de, 
  patrick@kph.uni-mainz.de, pochodza@kph.uni-mainz.de)}
}

\markboth{\bf 2008 IEEE Nuclear Science Symposium Conference Record}{\bf N17-7}
\maketitle

\begin{abstract}
  A tracking detector based on two meters long scintillating fibers
  read out by silicon photomultipliers (SiPM) is being developed for the \KAOS\
  spectrometer at the Mainz Microtron MAMI. Results from a prototype
  setup using 2\,mm square fibers and large area SiPM readout are presented.
  The detection efficiency of such a combination was measured to be between 83 and 100\%
  depending on the threshold on the SiPM amplitude. A Monte Carlo simulation based on
  a physical model was employed in order to extract the photon detection efficiency of the
  SiPM devices.
\end{abstract}


\section{Introduction}
\IEEEPARstart{T}{he}
recently upgraded electron accelerator MAMI-C with beam energies up to
1.5\,GeV has opened the door to kaon production experiments
at the Institut f\"ur Kernphysik in Mainz, Germany~\cite{Kaiser2008}.
The short orbit spectrometer \KAOS\
has been added to the existing facilities allowing the detection of short living kaons 
with a high survival probability. The simultaneous detection of scattered electrons and
positive kaons with this instrument at very forward angles will permit 
spectroscopic studies of hypernuclei by missing mass
reconstruction. The spectrometer was successfully operated at GSI near Darmstadt in heavy
ion collision experiments. In order to cope with the planned
experiments at MAMI the existing detector system has to be completed with a package for the electron
momentum and track reconstruction.  Timing and position information
can be obtained simultaneously by scintillating fiber tracking
detectors.  Multianode photomultipliers (MaPMT) have been used as 
readout devices for the vertical component of a detector of this type.
Experience has shown that there are several drawbacks associated with
MaPMT. In particular, optical cross-talk among neighboring channels has been observed
giving rise to a reduced position resolution~\cite{Achenbach2008}. In addition the need for
high voltage supplies and magnetic shielding increases the overall
price and complexity of the detector~\cite{Achenbach-SNIC06}.

\begin{figure}
  \centering
  \includegraphics[height=0.8\columnwidth]{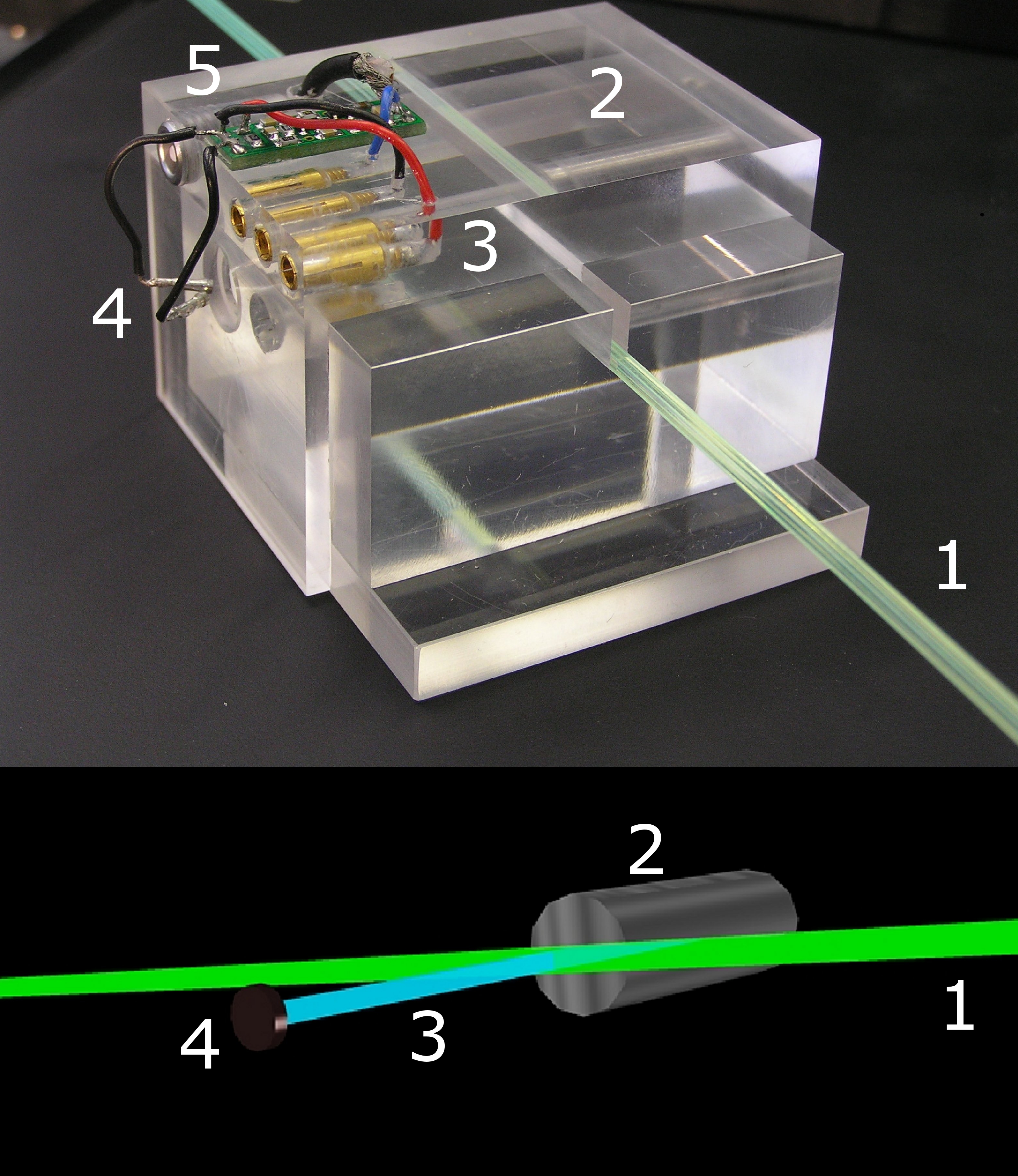}
  \caption{Modular Plexiglas device used for 
    measurements of the SiPM$/$fiber performance. 
    The two meters long green emitting fiber (1) is placed
    in a slit and kept in place by the top Plexiglas cover. A lead
    collimator (2) containing an $^{90}$Sr beta source with an opening of
    0.7\,mm is inserted in the central block perpendicular to the
    square fiber in good alignment with its center. The fraction of
    electrons that cross the first fiber penetrate a
    short blue emitting fiber (3) collinear with the collimator
    axis. The signal from this trigger fiber is read out by a blue
    sensitivity SiPM (4) and amplified by a transimpedance integrated
    amplifier (5).}
  \label{fig:Dcollisource}
\end{figure}

Moreover, a light sensor for the horizontal component of the tracking
detector has to be capable of a reliable operation in vacuum.
Silicon photomultipliers (SiPM) are emerging as a solid state
alternative to conventional photomultipliers in many fields. Hundreds
of micrometric avalanche photodiodes (APD) connected in parallel are
operated in a SiPM beyond the breakdown voltage for a high gain of
10${^6}$~\cite{Buzhan2001}.

\IEEEpubidadjcol
\section{Scintillating Fibers with SiPM Readout as Tracking Detectors}
SiPM have been suggested as possible 
readout device for the 150 two meters long scintillating fibers forming the
horizontal detector. Magnetic field insensitivity, small volume,
sensor independence, good vacuum performance and low voltage operation
make SiPM interesting for this application. Low light level detection
is on the other hand challenging for these devices due to their high
dark count rate. Long and thin fibers will only be capable of guiding
a few photons to the detector surface due to the low energy deposition
of minimum ionizing electrons and the strong light
absorption. Detection efficiency is a major issue for a tracking
system and the right combination of fiber and SiPM has to be carefully
studied.  Radiation hardness studies have also been performed showing
that commercially available SiPM suffer from a large increase in leakage
current for relatively low radiation doses~\cite{Sanchez2009}. This poses an additional
problem for their use in combination with long scintillator fibers. 
The leakage current will appear as a high rate of single photoelectron signals
due to the avalanche amplification specific of photodiodes operated in
the limited Geiger mode.

\pagestyle{plain}

\begin{figure}
  \centering
  \includegraphics[height=\columnwidth,angle=90]{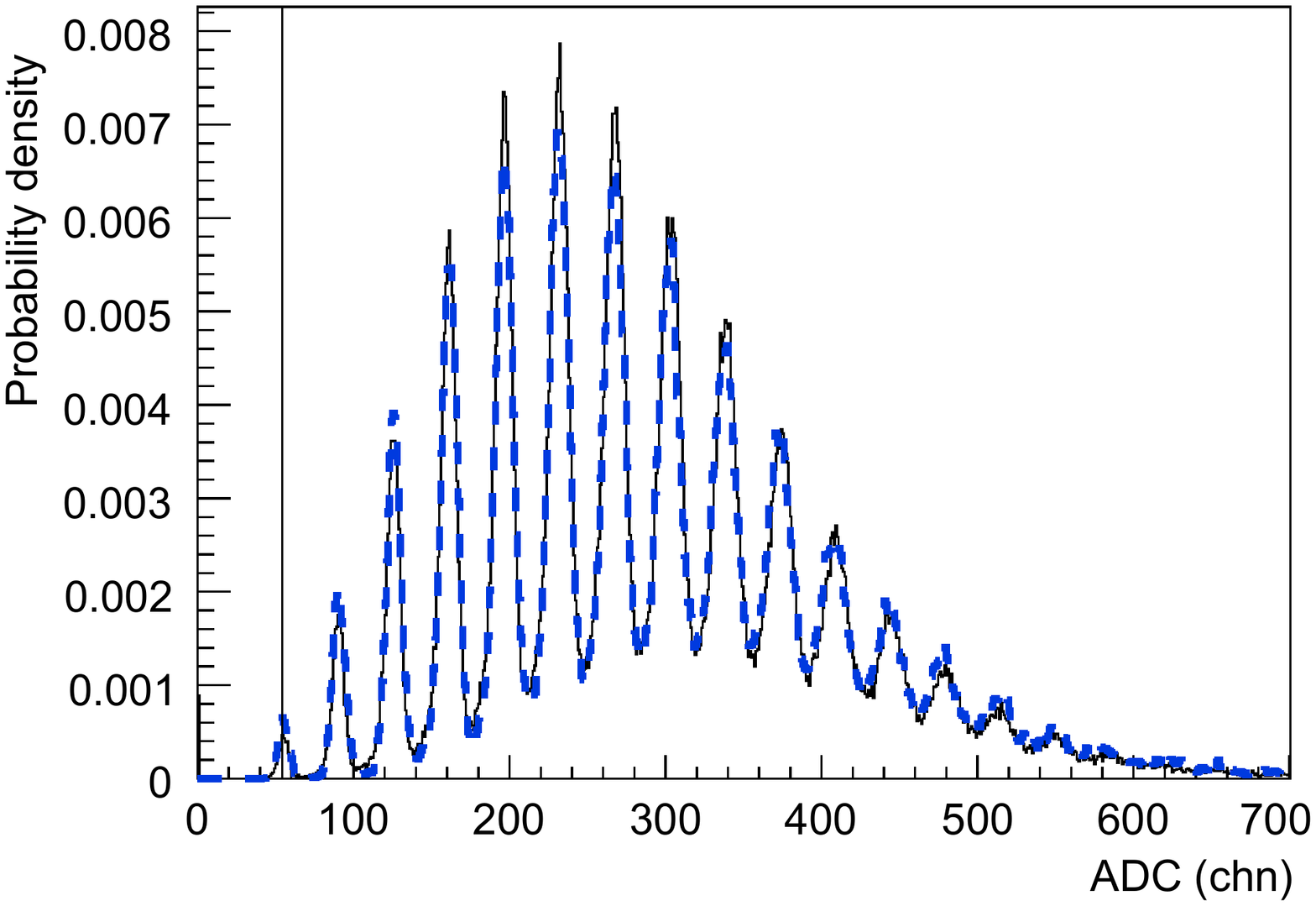}
  \caption{ADC spectrum for a Photonique
    device of 1\,mm$^2$ cross-section with type number SSPM-0701BG-TO18 
    illuminated by a low intensity light source. The position
      of the pedestal peak is indicated by a vertical line, the
      following peaks resolve the signals from single and multiple
      pixels of the SiPM. The peak structure is due to the narrow response 
      function. The bold-dashed curve is
    	the result of a Monte Carlo simulation including the main operating
    	parameters of SiPM.}
  \label{fig:D189fit}
\end{figure}

Fig.~\ref{fig:D189fit} shows a
typical ADC spectrum recorded for a 1\,mm$^{2}$ device from Photonique\footnote{Photonique SA, 
http://www.photonique.ch (2008)}, type number
SSPM-0701BG-TO18. The bold-dashed curve is the result of a Monte Carlo
simulation including the most relevant parameters of SiPM as optical
cross-talk, after-pulsing, photon detection efficiency (PDE) and gain variations developed in order
to extract the mean number of detected photons. This method is
necessary due to the overestimation that is quoted in many
publications where the effect of optical cross-talk and after-pulsing
is not subtracted. SiPM are manufactured so that signal uniformity
from pixel to pixel is quite good, typically within 10\%.
The small gain variation together with the narrow single electron response
function of each APD provides excellent photon counting capabilities
as can be appreciated in the well defined peak structure of the
spectrum.

It is well known that 
SiPM noise rate is mostly due to single pixel signal.
Rates of the order of several megahertz at room temperature are normal
in todays commercially available SiPM.  An avalanche of 10$^6$
carriers in any of the micrometric APD forming the SiPM will create
around 50 photons via hot carrier luminescence with enough energy to
trigger any neighboring pixel~\cite{Otte2006}. This phenomenon is
known as optical cross-talk and explains the measurable dark rate for
threshold beyond one pixel signal amplitude.  These signals compete
with real signals generated by a small number of photons.  
A simple model based on the probability $q$ of single
neighbor activation allows the probabilities for the different
clusters of APD to be calculated (only pixels sharing one complete
side are allowed as members of a cluster). Cluster probabilities are
given by the zero pixel cross-talk probability $P(0)=(1-q)^4$, and the
$N$-pixel cross-talk probabilities $P(1)=4q(1-q)^6$,
$P(2)=q^2[6(1-q)^8+12(1-q)^7]$, and
$P(3)=q^3[32(1-q)^8+32(1-q)^9+8(1-q)^{10}]$.

\begin{figure}
  \centering
  \includegraphics[height=\columnwidth,angle=90]{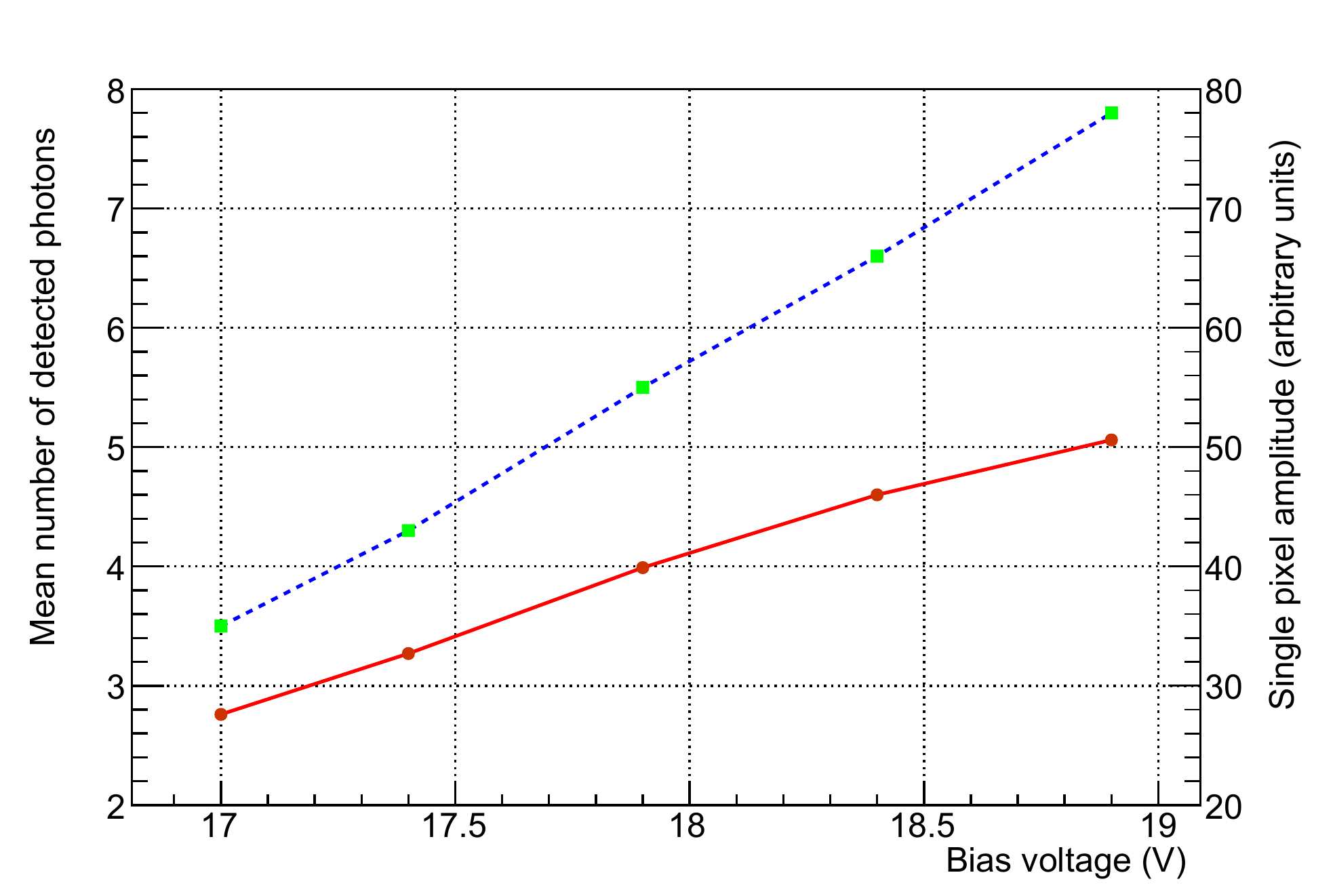}
  \caption{Mean number of detected photons and single pixel amplitude
    as a function of the bias voltage for two Photonique
    devices of 1\,mm$^2$ cross-section with type number SSPM-0701BG-TO18. The
    observed differences between the two curves are not fully
    understood but manufacturing variability seems to be the most
    probable explanation. The measured linear
    dependence is explained by the linear increase in avalanche probability and
    diode capacity charging.}
  \label{fig:DPDEvsvolt}
\end{figure}
\begin{figure}
  \centering
  \includegraphics[height=\columnwidth,angle=90]{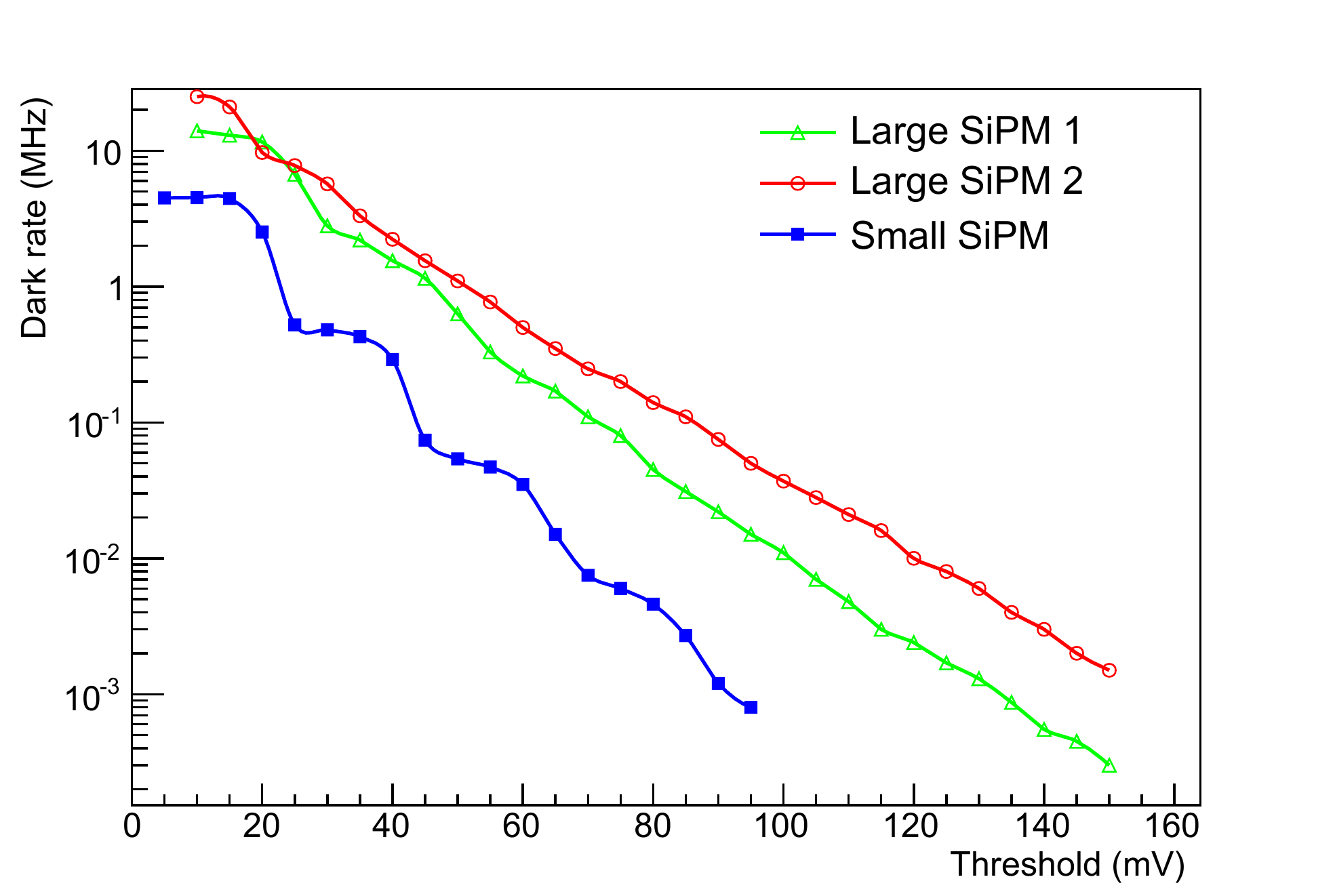}
  \caption{Dark count rate as a function of threshold level
    measured for two large area SiPM SSPM-0606BG4MM-PCB and for
    a 1 mm$^{2}$ device. The dark count rate is a factor of four higher
    for the larger devices and steps are less defined due to the
    more probable signal pile-up.}
  \label{fig:Descaleragrandepeq}
\end{figure}
\begin{figure}
  \centering
  \includegraphics[width=\columnwidth]{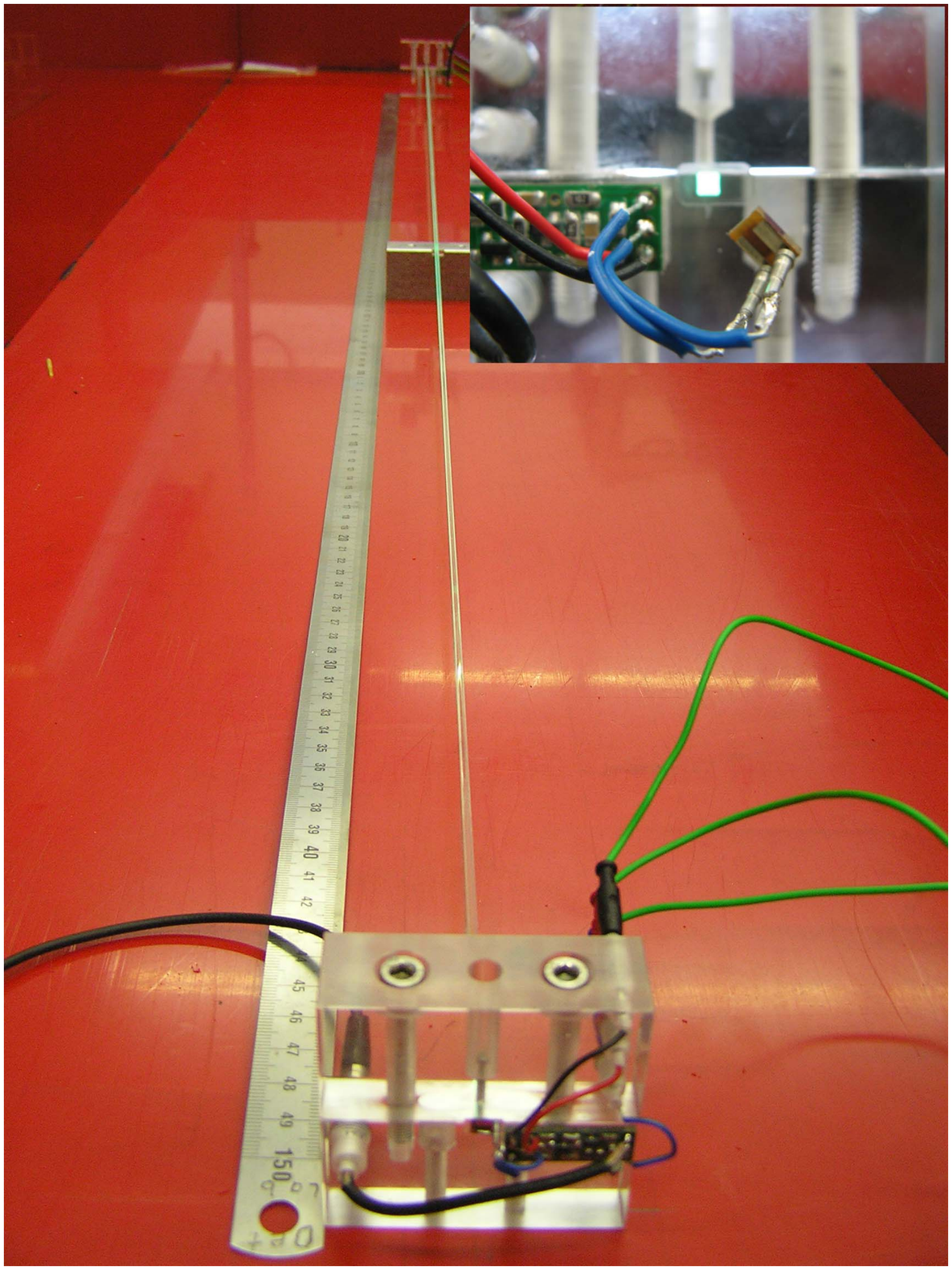}
  \caption{Test setup inside a light tight box containing the two meter long fiber under
    study. The inset shows the transparent connectors on both extremes that allow
    optimal alignment of SiPM and fiber. No specific optical coupling
    was necessary due to the good matching of indexes of refraction of
    the protective epoxy layer of the SiPM and the fiber
    core material.}
  \label{fig:Dcaja}
\end{figure}

\section{Measurements on SiPM Performance}
Previous studies performed with 0.85\,mm diameter 2\,m long cylindrical
fibers read out by Photonique SiPM with an active area of 1\,mm$^{2}$
showed that the small number of generated photons was a serious
concern for efficient electron detection.  
Fig.~\ref{fig:DPDEvsvolt} shows the mean ADC value as
function of the applied voltage for this SiPM. The almost liner dependency allows to conclude that low
voltages are more appropriate for low light level detection due the
much faster increase (exponential) in the dark count rate with
voltage. The differences in the measurements for the two devices are not
fully understood although manufacturing variability seems to be the
most probable explanation. A larger energy deposition
in the scintillating material is necessary in order to achieve
detection efficiencies close to 100\%.  On the other hand a minimum particle
trajectory disturbance is important for a tracking detector.  

2\,mm square fibers (emission peak at 492\,nm) with double cladding from Bicron\footnote{Bicron, 
http://www.bicron.com (2008)} 
with type number BCF-20 were
considered an adequate compromise for the application.  
For the chosen fiber the manufacturer quoted a value of 7.3\% trapping efficiency. 
For the SSPM-0606BG4MM-PCB device
typical PDE values range from 15 to 25\% with the PDE almost
constant over the wavelength range 500--650\,nm. 
However, the PDE is linear function of the bias voltage due to the increase in probability
of avalanche initiation.  The typical bias voltage
for these devices is 30\,V.
The maximum number of detected 
photons from such a combination can be estimated with the above values to be of the order of 12 photons. 

\begin{table*}
  \caption{Measured particle detection efficiencies for 2\,mm square fibers 
    and 0.86\,mm cylindrical fibers being read out by SiPM as a 
    function of threshold value in units of single pixel amplitude. The random coincidence rate
    was determined from dark count rate measurements.}
  \centering
  \renewcommand*{\arraystretch}{2}
  \renewcommand*{\tabcolsep}{0.4cm}
  \begin{tabular}{|c|c|c|c|c|}
    \hline
    	  & \multicolumn{2}{c|}{\normalsize 2\,mm Fiber $/$ 4.4\,mm$^2$ SiPM}
    		& \multicolumn{2}{c|}{\normalsize 0.86\,mm Fiber $/$ 1\,mm$^2$ SiPM}\\ \hline
    Threshold  (pixel) & Efficiency (\%) & Random Rate (kHz)  & Efficiency (\%) & Random Rate (kHz)\\  
    \hline\hline
    \normalsize   0.5& \normalsize  100 & \normalsize 2\,000 & \normalsize  91 & \normalsize 320  \\ \hline
    \normalsize   1.5& \normalsize  99.8& \normalsize   80 & \normalsize  76 & \normalsize   5  \\ \hline
    \normalsize   2.5& \normalsize  95.0& \normalsize  1.3 & \normalsize  56 & \normalsize 0.45 \\ \hline
    \normalsize   3.5& \normalsize  82.6& \normalsize 0.04 & \normalsize  35 & \normalsize 0.04 \\ \hline
  \end{tabular}
  \label{fig:table}
\end{table*}

The measured dark count rate as a
function of the threshold level in a leading edge discriminator is shown
for the two 4.4\,mm${^2}$ and for a 1\,mm${^2}$ device in Fig.~\ref{fig:Descaleragrandepeq}.
The step like structure 
can be explained by the cross-talk probabilities discussed above. For large area devices steps
are much less defined due to signal pile-up. It is clear from this
measurement that the threshold level is a crucial issue for noise
reduction. Fibers of 4\,mm${^2}$ cross-section will only increase the amount of
generated light by a factor of 2 but noise in the SiPM will be 4 times
larger due to the fact that dark count rate increases linearly with
the detector surface. The improvement comes from the fact that high
efficiencies can be achieved with higher thresholds for the larger
cross-section fibers and the geometrical reduction of dark count rate
with threshold level will allow an effectively lower
rate. 

\begin{figure}
  \centering
  \includegraphics[height=\columnwidth,angle=90]{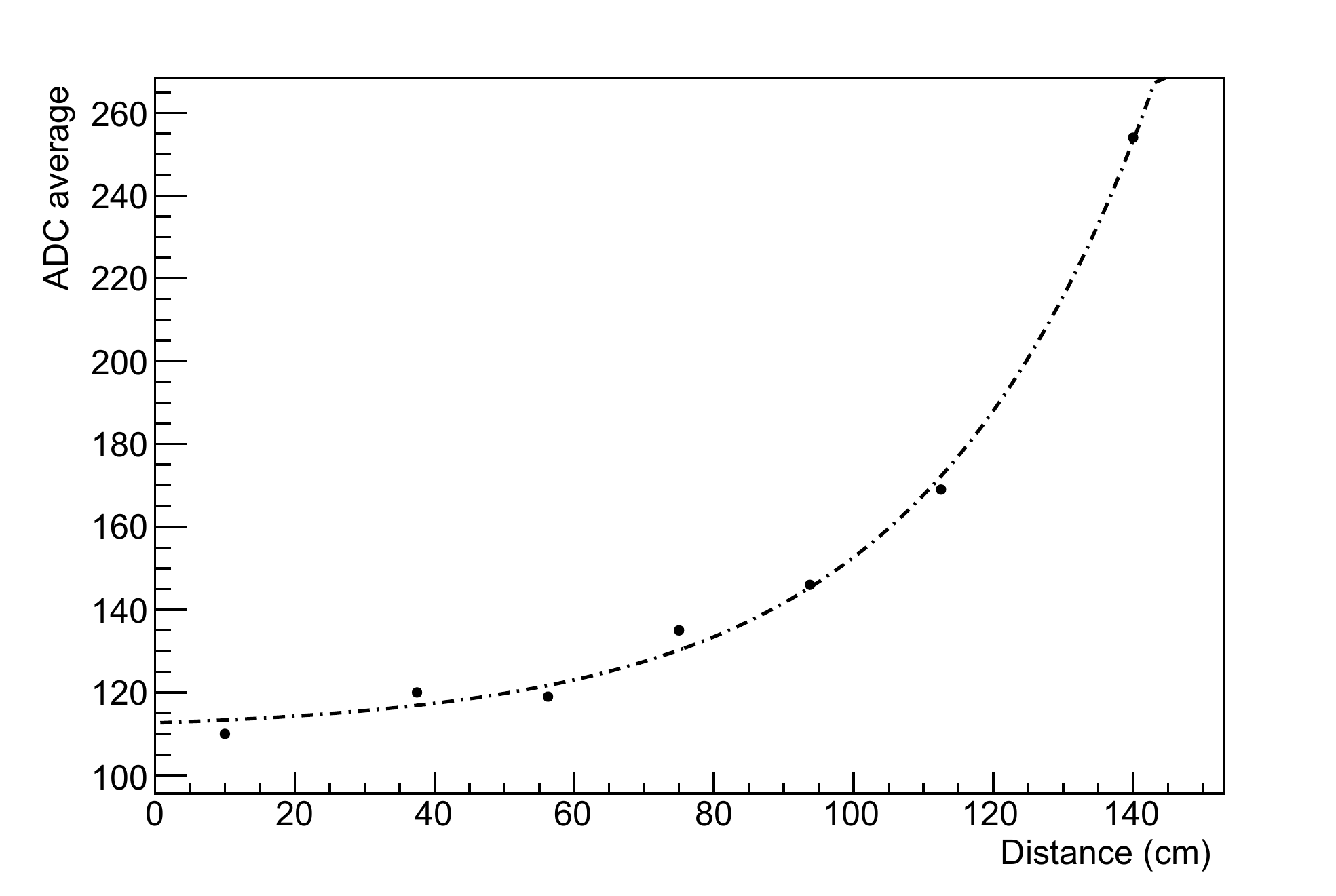}
  \caption{Mean ADC value measured as a function of the beta source
    position. A characteristic exponential dependence is observed but
    the attenuation length seems to substantially smaller than the value
    quoted by the manufacturer.}
  \label{fig:Dabsorption}
\end{figure}

Fig.~\ref{fig:Dcaja} shows the setup for characterizing 
a two meters long Bicron BCF-20 fiber
read out in both extremes by the large area SiPM. Signals are brought into a compact
electronic board incorporating a SiPM bias circuit and a
transimpedance amplifier optimized for the amplification of SSPM
signals. The coupling to the scintillating fiber is direct. No
optical connection is necessary due to the small difference in
the refraction indexes of the protecting epoxy layer used for the SiPM
and the fiber core material. A Plexiglas connector was designed that allowed a
reliable connection and mechanical stability of the full assembly as
well as optical control of the relative position of fiber to SiPM for a
proper alignment. A 250\,cm long light tight box was constructed to
keep the experimental arrangement protected from external light
sources. The voltage for the preamplifiers and the SiPM was constantly monitored
and the temperature was measured to be stable within 2\,$^\circ$C.  Light
absorption was measured by exciting the fiber at several points with a
beta source and by determining the corresponding change in the mean ADC
value, see Fig.~\ref{fig:Dabsorption}. The measured absorption length of 1.5\,m 
is substantially smaller than the value quoted by the manufacturer
($>$ 3.5\,m for 1\,mm diameter fiber measured with a bialkali cathode PMT)
which needs clarification.  

\begin{figure}
  \centering
  \includegraphics[width=\columnwidth]{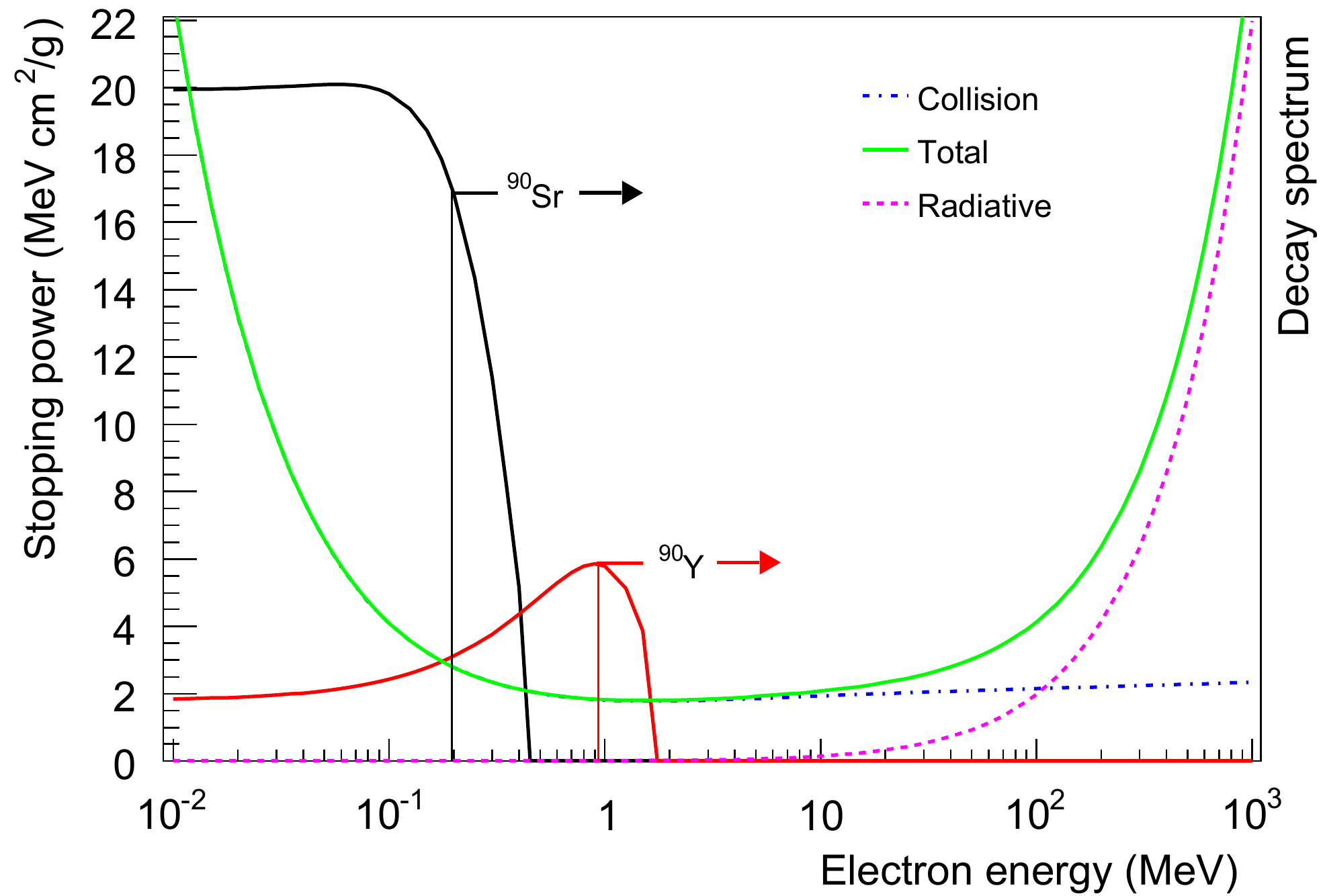}
  \caption{Stopping power of organic fibers for electrons separated into collision, radiative 
  	and total losses as a function of energy. The decay spectra of $^{90}$Sr 
  	and $^{90}$Y and the mean kinetic energy of the decay electrons were overlaid onto this
  	plot.}
  \label{fig:Dstopingpower}
\end{figure}

A simple calculation shows the equivalence of the energy loss in scintillating fiber
tests made with the $^{90}$Sr source to the loss of high energy
electrons. $^{90}$Sr is an beta-source with a maximum
kinetic energy of $E_{max}=$ 546\,keV and a mean kinetic
energy of $E=$ 196\,keV. It decays to $^{90}$Y with a half-life of
29.12\,y. The short-lived daughter decays with a maximum
kinetic energy of $E_{max}=$ 2.28\,MeV and a mean energy of $E=$ 933\,keV.
The spectra of both decays are shown in Fig.~\ref{fig:Dstopingpower}.  
Most electrons from the $^{90}$Sr decay are stopped 
in a 2\,mm thick scintillator, since the range of 0.35\,MeV electrons in
scintillator material is only $R=$ 1\,mm.
It is readily seen from the stopping power curves that ionizing energy loss of the
$\sim$ 1\,MeV electrons from $^{90}$Y decay will be
equivalent to that of high energy electrons.  

The efficiency was
measured with a dedicated device consisting of a 3\,cm long cylindrical
lead collimator with an aperture of 0.7\,mm in diameter and containing a
$^{90}$Sr beta source, see Fig.~\ref{fig:Dcollisource}. Electrons 
were forced to cross the 2\,m long
fiber by a Plexiglas structure with a slit to
accommodate the studied fiber. A 2\,cm long blue scintillating fiber
was introduced in the Plexiglas parallel to the electron trajectory so
that those that are able to go through the studied fiber will enter
longitudinally the small one depositing all their remaining energy
there. The relatively large amount of generated photons is detected by
a blue sensitive Photonique device with type number SSPM-0611B1MM-TO18. The signal
amplifier is located in the Plexiglas block as well so that a very
compact device was obtained that could be freely moved along the fiber.
Practically no absorption takes place in the 2\,cm short trigger fiber
and high thresholds for the trigger discriminator could be chosen. The
background rate in absence of the exciting source was almost zero.  
Efficiency was defined as
the quotient of trigger signals to coincidences of the left and right
large area SiPM. This value was studied as a function of the threshold
level in units of single pixel signals. The results for the
cylindrical and square fibers are shown in Table~\ref{fig:table}.

\enlargethispage{-1cm}

\section{Conclusion}
The \KAOS\ spectrometer at MAMI will be extended by a large fiber detector in the near future. Our study has shown that the readout of 2\,mm square scintillating fibers by SiPM can lead to near 100\% detection efficiency for electrons. A random coincidence rate for a setup of two 4.4\,mm$^2$ large area SiPM is unavoidable because of high dark count rate of these devices. However, with a reasonable threshold setting and further trigger conditions such a system is feasable.

\bibliographystyle{IEEEtran}
\bibliography{KAOS-28-07-08}

\end{document}